\documentclass[11pt]{article}
\usepackage{graphicx}
\usepackage{amssymb}

\newtheorem{theorem}{Theorem}

\newtheorem{lemma}{Lemma}

\def\proof{\noindent  {\underline {Proof}}. }
\def\square{ {\hfill \vrule height6pt width6pt depth1pt} \bigskip \medskip }

\begin{document}

\centerline{\Large Riemann surfaces, separation of variables and classical}
\centerline{\Large  and quantum integrability.}
\vskip 1cm
\centerline{O. Babelon $^{1} $  and M. Talon
\footnote{Member of CNRS.}}
\vskip .5cm
\centerline{Laboratoire de Physique Th\'eorique et Hautes Energies.
\footnote{L.P.T.H.E. Universit\'es Paris VI--Paris VII (UMR 7589),
 Bo\^{\i}te 126, Tour 16, $1^{er}$ \'etage,
 4 place Jussieu, F-75252 PARIS CEDEX 05}
}
\date{September 2002}

\vskip 5cm

\begin{abstract}
We show that Riemann surfaces, and separated variables immediately 
provide classical Poisson commuting Hamiltonians.
We show that Baxter's equations for separated variables immediately 
provide quantum commuting Hamiltonians.
The construction is simple, general, and does not rely on the 
Yang--Baxter equation.
\end{abstract}

\vfill
PAR LPTHE 02--43 
\eject

\section{Introduction.}
We know since Liouville that integrability means commuting Hamiltonians. 
It is the primary role of Lax matrices and the Yang-Baxter equation to provide 
non trivial such Hamiltonians.
In the classical theory, additional benefits are the spectral curve $\Gamma$
 and the ability to separate variables by considering $g  = {\rm genus } (\Gamma)$ points
on it \cite{DuKrNo90}. \\
In the quantum theory, the analog construction is Sklyanin's method of separation of variables 
and Baxter's equations \cite{Skly92,Skly95}.  Despite the beauty of this result, 
the route  from a Yang--Baxter defined quantum integrable model 
to the separated variables is usually  long and difficult, especially in the non hyperelliptic case.\\
Here, we show that we can reverse the strategy. We start from separated variables and 
consider Baxter's equations as equations for the Hamiltonians.
We then prove that these Hamiltonians commute under very general
hypothesis.  \\
By its generality, its simplicity and its close analogy to the classical case,  
this result could provide a good starting 
point to build a theory of quantum integrable systems.

\section{The main theorem.}

Consider a curve in $C^2$
\begin{equation}
\Gamma (\lambda,\mu) \equiv R_0(\lambda,\mu) + \sum_{j=1}^g R_j(\lambda,\mu) H_j =0
\label{rlambdamu}
\end{equation}
where the $H_i$ are the only dynamical moduli,  so that $R_0(\lambda,\mu)$ and $R_i(\lambda,\mu)$ do not contain any dynamical variables. 
If things are set up so that $\Gamma$ is of genus $g$ and there are exactly $g$ Hamiltonian
$H_j$ (see below for  realizations of this setup), then the curve is completely determined 
by requiring that it passes through $g$ points $(\lambda_i, \mu_i)$, $i=1,\cdots, g$. Indeed, the  moduli $H_j$ are determined by solving the linear system
\begin{equation}
 \sum_{j=1}^g R_j(\lambda_i,\mu_i) H_j + R_0(\lambda_i,\mu_i) = 0,\quad i=1,\cdots ,g
\label{baxterclassique}
\end{equation}
whose solution is 
\begin{equation}
 H = - B^{-1} V
\label{hamiltoniensclassiques}
\end{equation}
where
$$
H = \pmatrix{H_1 \cr \vdots \cr H_i \cr \vdots \cr H_g},\quad
B = \pmatrix{ 
 R_1(\lambda_1,\mu_1) & \cdots & R_g(\lambda_1,\mu_1) \cr
\vdots   &     &  \vdots  \cr
 R_1(\lambda_i,\mu_i) & \cdots & R_g(\lambda_i,\mu_i) \cr
\vdots   &     &  \vdots  \cr
 R_1(\lambda_g,\mu_g) & \cdots & R_g(\lambda_g,\mu_g) },\quad
V = \pmatrix{ R_0(\lambda_1,\mu_1) \cr \vdots \cr  R_0(\lambda_i,\mu_i)
\cr \vdots \cr  R_0(\lambda_g,\mu_g)  }
$$
Here, of course, we assume that generically $\det B \neq 0$.

\begin{theorem}
Suppose that the variables $(\lambda_i, \mu_i)$ are separated i.e. they Poisson commute for 
$i \neq j$:
\begin{equation}
\{ \lambda_i,\lambda_j\} = 0, \quad \{ \mu_i,\mu_j\}=0, \quad
\{ \lambda_i,\mu_j\} = p(\lambda_i, \mu_i) \delta_{ij}
\label{poisep}
\end{equation}
Then the Hamiltonians $H_i$, $i=1\cdots g$, defined by eq.(\ref{hamiltoniensclassiques}) Poisson commute
$$
\{ H_i , H_j \} = 0
$$
\end{theorem} 
\proof
Let us compute
\begin{eqnarray*}
B_1 B_2 \{ (B^{-1}V)_1,(B^{-1}V)_2 \} &=& 
\{ B_1, B_2 \}  (B^{-1}V)_1 (B^{-1}V)_2 \\
&& \hskip -2cm- \{ B_1, V_2 \}  (B^{-1}V)_1
- \{ V_1, B_2 \}  (B^{-1}V)_2
+  \{ V_1, V_2 \}
\end{eqnarray*}
Taking the matrix element $i,j$ of this expression, we get
\begin{eqnarray*}
\Big( B_1 B_2 \{ (B^{-1}V)_1,(B^{-1}V)_2 \}\Big)_{ij} &=& 
\delta_{ij}\sum_{k,l} \{ B_{ik}, B_{il} \}  (B^{-1}V)_k (B^{-1}V)_l \\
&& \hskip -5cm- \delta_{ij}\sum_k\{ B_{ik}, V_i \}  (B^{-1}V)_k
- \delta_{ij}\sum_l \{ V_i, B_{il} \}  (B^{-1}V)_l
+  \delta_{ij}\{ V_i, V_i \} =0
\end{eqnarray*}
where $\delta_{ij}$ occurs because the variables are separated.
\square

It can hardly be simpler. The only thing we use is that the Poisson bracket
vanishes between different lines of the matrices, and then the antisymmetry.
We did not even need to specify the Poisson bracket  between $\lambda_i$ and 
$\mu_i$. The Hamiltonian are in involution whatever this Poisson bracket is.
This is the root of the multihamiltonian structure of integrable systems.

Can we make it quantum? Let us consider a set of separated variables
$$
\relax [ \lambda_i,\lambda_j ] = 0, \quad [ \mu_i,\mu_j] =0, \quad
[ \lambda_i,\mu_j]  = p(\lambda_i, \mu_i) \delta_{ij}
$$
We want Baxter's equation, so we start 
from the linear system
\begin{equation}
\sum_j R_j(\lambda_i,\mu_i) H_j + R_0(\lambda_i,\mu_i) = 0
\label{baxter}
\end{equation}
Here the $ H_j$ are on the right, and in $R_j(\lambda_i,\mu_i)$, 
$R_0(\lambda_i,\mu_i)$, we assume some order between $\lambda_i,\mu_i$,
but the coefficients in these functions are non dynamical. Hence we start from the linear system
\begin{equation}
B H = -V
\label{baxtercompact}
\end{equation}
We notice that we can define unambiguously the left inverse of $B$.
First, the determinant $D$ of $B$ is well defined because it never 
involves a product of elements on the same line. The same is true
for the cofactor $\Delta_{ij}$ of the element  $B_{ij}$ (we include the sign $(-1)^{i+j}$ in the definition of $\Delta_{ij}$). 
Define
$$
B^{-1}_{ij} \equiv (B^{-1})_{ij} = D^{-1} \Delta_{ji}
$$
We have 
$$
(B^{-1} B)_{ij} = \sum_k D^{-1} \Delta_{ki} B_{kj}
$$
But $\Delta_{ki}$ does {\em not} contain any element $B_{kl}$, hence
the product $ \Delta_{ki} B_{kj}$ is commutative, and the usual 
construction of the inverse of $B$ is still valid. Since the left and right
inverse coincide in an associative algebra with unit, we have the identities
\begin{equation}
(BB^{-1})_{ij}=\sum_k  B_{ik} B^{-1}_{kj}
=\sum_k B_{ik} D^{-1} \Delta_{jk} = \delta_{ij}
\label{identityone}
\end{equation}
We write the solution of eq.(\ref{baxtercompact}) as
\begin{equation}
H = - B^{-1} V
\label{commutingquantities}
\end{equation}

\begin{theorem}
The quantities $H_i$ defined by eq.(\ref{commutingquantities}), which
solve Baxter's equations eqs.(\ref{baxter}),
are all commuting
$$
\relax [ H_i , H_j ] = 0
$$
\end{theorem}
\proof
Using that $V_k$ and $V_l$ commute, $[ V_k,V_l ] = 0$, we compute 
\begin{eqnarray}
\relax [ H_i, H_j ] &=& \sum_{k,l}\; [ B^{-1}_{ik} V_k , B^{-1}_{jl} V_l ] \label{cicj} \\
&=&\sum_{k,l}\;  [B^{-1}_{ik}, B^{-1}_{jl} ] V_k V_l 
- B^{-1}_{ik} [B^{-1}_{jl}, V_k ] V_l 
+ B^{-1}_{jl} [B^{-1}_{ik}, V_l ] V_k \nonumber
\end{eqnarray}
Using 
$$
\relax [A^{-1}, B^{-1} ] =A^{-1} B^{-1}[A,B]B^{-1}A^{-1}= 
B^{-1} A^{-1}[A,B]A^{-1}B^{-1}
$$
so that
\begin{eqnarray*}
\relax [ B_{ik}^{-1}, B_{jl}^{-1} ] &=&  \sum_{rs,r's'}B^{-1}_{ir}B^{-1}_{jr'}
[B_{rs}, B_{r's'}] B^{-1}_{s'l}B^{-1}_{sk} \\
&=&  \sum_{rs,r's'}B^{-1}_{jr'} B^{-1}_{ir}
[B_{rs}, B_{r's'}]B^{-1}_{sk} B^{-1}_{s'l}
\end{eqnarray*}
the first term can be written
\begin{eqnarray*}
\sum_{k,l}\; [B^{-1}_{ik}, B^{-1}_{jl} ]  V_kV_l &=&
\sum {1\over 2} B^{-1}_{ir}B^{-1}_{jr'}  [B_{rs}, B_{r's'}]
 \Big( B^{-1}_{s'l} B^{-1}_{sk} + B^{-1}_{s'k}  B^{-1}_{sl} \Big) V_kV_l \\
&=&\sum {1\over 2}B^{-1}_{jr'} B^{-1}_{ir}  [B_{rs}, B_{r's'}]
 \Big(  B^{-1}_{sk}B^{-1}_{s'l} +  B^{-1}_{sl} B^{-1}_{s'k} \Big) V_kV_l \\
\end{eqnarray*}
Using that $[B_{rs}, B_{r's'}] = \delta_{rr'}[B_{rs}, B_{rs'}]$ and is therefore
antisymmetric in $ss'$, and setting
$$
K_{ss'} = \sum_{k,l}\; \Big( 
B^{-1}_{s'l} B^{-1}_{sk} + B^{-1}_{s'k}  B^{-1}_{sl} 
-B^{-1}_{sl} B^{-1}_{s'k} - B^{-1}_{sk}  B^{-1}_{s'l} 
\Big) V_kV_l
$$
 we get
\begin{eqnarray*}
\sum_{k,l} [B^{-1}_{ik}, B^{-1}_{jl} ]  V_kV_l &=&
\sum_{rss'} {1\over 4} B^{-1}_{ir}B^{-1}_{jr}  [B_{rs}, B_{rs'}] K_{ss'} \\
&=&-\sum_{rss'} {1\over 4}B^{-1}_{jr} B^{-1}_{ir}  [B_{rs}, B_{rs'}]K_{ss'} \\
&=&\sum_{rss'} {1\over 8}[B^{-1}_{ir}, B^{-1}_{jr}]  [B_{rs}, B_{rs'}]K_{ss'} 
 \end{eqnarray*}
The last two terms in eq.(\ref{cicj}) are simpler, we get
\begin{eqnarray*}
\sum_{k,l}\;  B^{-1}_{jl} [B^{-1}_{ik}, V_l ] V_k
 - B^{-1}_{ik} [B^{-1}_{jl}, V_k ] V_l =
\sum_{rsk} [B^{-1}_{ir}, B^{-1}_{jr}]  [B_{rs}, V_{r}] B_{sk}^{-1} V_k
\end{eqnarray*}
The quantities $H_i$ will commute if
\begin{equation}
 [B^{-1}_{ir}, B^{-1}_{jr}] = 0, \quad \forall i,j,r
\label{comb-1}
\end{equation}
This is true as shown in the next Lemma. \square

The condition eq.(\ref{comb-1}) says that the elements on the same column of 
$B^{-1}$  commute among
themselves. In a sense this is a condition dual to the one on $B$. It is true 
semiclassically because
$$
\{ B^{-1}_{ir}, B^{-1}_{jr} \} = \sum_{a,a',b,b'} B^{-1}_{ia} B^{-1}_{ja'}
\{ B_{ab}, B_{a'b'} \} B^{-1}_{br} B^{-1}_{b'r} =
\sum_{a,b,b'} B^{-1}_{ia} B^{-1}_{ja}
\{ B_{ab}, B_{ab'} \} B^{-1}_{br} B^{-1}_{b'r}=0
$$
where in the last step we use the antisymmetry of the Poisson bracket.
We show that it is also true quantum mechanically\footnote{The same Lemma already appeared in  \cite{ER01}. Our two proofs are different and independent however.   We thank B. Enriquez for drawing our attention to his work.}.
\begin{lemma}
Let $B$ be a matrix whose elements commute if they do not belong 
to the same line
$$
\relax [ B_{ik} , B_{jl} ] = 0 \quad {\rm if } \; i \neq j
$$
Then the inverse $B^{-1}$ of $B$ is defined without ambiguity and
moreover elements on a same column of $B^{-1}$ commute
$$
\relax  [ B^{-1}_{ir} , B^{-1}_{jr} ] = 0 
$$
\end{lemma}
\proof
We want to show that
$$
\Delta_{ri}B^{-1}_{jr} = \Delta_{rj}B^{-1}_{ir}
$$
denote by $\beta_i^{(r)}$ the vector with components $B_{ki}$, $k \neq r$. Then
we have (with $j>i$)
\begin{eqnarray*}
\Delta_{ri}B^{-1}_{jr} &=& (-1)^{r+i}
 \beta^{(r)}_1 \wedge \beta^{(r)}_2 \wedge \cdots \widehat{ \beta^{(r)}_i} 
\wedge \cdots \beta^{(r)}_j \wedge \cdots\beta^{(r)}_g 
B^{-1}_{jr}\\
&=& (-1)^{r+i+g-j}\beta^{(r)}_1 \wedge \beta^{(r)}_2 \wedge \cdots \widehat{ \beta^{(r)}_i} 
\wedge \cdots \widehat{\beta^{(r)}_j} \wedge \cdots\beta^{(r)}_g \wedge
\beta^{(r)}_j B^{-1}_{jr} \\
&=& (-1)^{r+i+g-j+1}\beta^{(r)}_1 \wedge \beta^{(r)}_2 \wedge \cdots \widehat{ \beta^{(r)}_i} 
\wedge \cdots \widehat{\beta^{(r)}_j} \wedge \cdots\beta^{(r)}_g \wedge
\sum_{k\neq j}\beta^{(r)}_k B^{-1}_{kr} \\
&=& (-1)^{r+i+g-j+1}\beta^{(r)}_1 \wedge \beta^{(r)}_2 \wedge \cdots \widehat{ \beta^{(r)}_i} 
\wedge \cdots \widehat{\beta^{(r)}_j} \wedge \cdots\beta^{(r)}_g \wedge
\beta^{(r)}_i B^{-1}_{ir} \\
&=&(-1)^{r+j}\beta^{(r)}_1 \wedge \beta^{(r)}_2 \wedge \cdots  \beta^{(r)}_i
\wedge \cdots \widehat{\beta^{(r)}_j} \wedge \cdots\beta^{(r)}_g 
B^{-1}_{ir} \\
&=& \Delta_{rj}B^{-1}_{ir}
\end{eqnarray*}
where in the third line we used eq.(\ref{identityone}). In the above manipulations, we never have two operators $B_{ij}$ on the same line
so we can use the usual properties of the wedge product. Moreover it is important that
the line $r$ is absent in the definition of $\beta^{(r)}$.
Remark that this equation can also be written 
$ \Delta_{ri}D^{-1}  \Delta_{rj}=  \Delta_{rj}D^{-1}  \Delta_{ri}$ which is a Yang--Baxter type equation.
\square

With this Lemma, we have completed the proof of our theorem. 
It is remarkable that, again, only the separated nature of the variables $\lambda_i, \mu_i$
is used in this construction, but the precise commutation relations between 
$\lambda_i, \mu_i$ does not even need to be specified. This is the origin of 
the multi Hamiltonian structure of integrable systems, here extended to the quantum
domain.

\section{Choosing the right number of dynamical moduli.}

Let us explain how one can set up things in order that 
the number of dynamical moduli is equal to the genus of the Riemann surface.
To understand the origin  of the conditions we will write, 
let us explain  first what happens in the setting of general rational Lax matrices
described in \cite{BaTa99,BaBeTa03}.
Quite generally, a Lax matrix $L(\lambda)$ depending 
 rationally on a spectral parameter $\lambda$, with poles at points $\lambda _k$
can be written as
\begin{eqnarray}
L(\lambda) = L_0 + \sum_k L_k(\lambda) 
\label{laxmatrix}
\end{eqnarray}
where $L_0={\rm Diag}(a_1,\cdots,a_N)$ is a constant diagonal matrix and
$L_k(\lambda)$ is
the polar part of $L(\lambda )$ at $\lambda _k$, ie. $L_k(\lambda)=\sum_{r=-n_k}^{-1}
L_{k,r} (\lambda -\lambda _k)^r$.  
In order to have a good phase space to work with, we  assume that $L_k(\lambda)$ 
lives in a coadjoint orbit of the group of $N\times N$ matrix regular in the vicinity of
$\lambda =\lambda _k$, i.e.
$$
L_k = \Big(g_k A_k g_k^{-1}\Big)_-
$$
Here  $A_k(\lambda )$ is a diagonal matrix with a pole of order
$n_k$ at $\lambda =\lambda _k$, and $g_k$ has a regular expansion at $\lambda =\lambda _k$.  The notation $()_-$ means
taking the singular part at $\lambda =\lambda _k$. This singular part only
depends on the singular part $(A_k)_-$ and the first $n_k$
coefficients of the expansion of $g_k$ in powers of $(\lambda -\lambda _k)$.  
The matrix $(A_k)_-$ is an orbit invariant which specifies the coadjoint orbit, and is
 {\em not} a dynamical variable. It is in the center of the Kirillov bracket which 
as shown in \cite{BaTa99} induces the Poisson bracket 
eq.(\ref{poisep}), with $p(\lambda_i,\mu_i)=1$, on the separated variables. 
The physical degrees of freedom are contained in the first $n_k$
coefficients of $g_k(\lambda )$. Note however that since $A_k$ commutes 
with diagonal
matrices one has to take the quotient by $g_k\to g_k d_k$
where $d_k(\lambda )$ is a regular diagonal matrix, in order to
correctly describe the dynamical variables on the orbit. 
The dimension of the orbit of $L_k$ is thus $N(N-1)n_k$ so that
$L(\lambda )$ depends on $\sum_k N(N-1)n_k$ degrees of freedom. 
Finally, the form and analyticity properties of
$L(\lambda )$ are invariant under conjugation by constant matrices. To
preserve the normalization, $L_0$, at $\infty$ these matrices have to be diagonal
(if all the $a_i$'s are different).   Generically, these
transformations  reduce the dimension of the 
phase space  by $2(N-1)$, yielding:
$$
{\rm dim}\,{\cal M}=(N^2-N)\sum_k n_k -2(N-1)
$$
The spectral curve is
\begin{eqnarray}
\Gamma~~ :~~ R(\lambda, \mu ) \equiv \det( L(\lambda )-\mu~{\bf 1} )
=(-\mu)^N +\sum_{q=0}^{N-1} 
r_q(\lambda) \mu^q =0
\label{specurve2} 
\end{eqnarray}
The coefficients $r_q(\lambda)$ are polynomials in the matrix elements
of $L(\lambda )$ and therefore have poles at $\lambda _k$. The curve is naturally presented as a $N$--sheeted covering of the $\lambda$-plane. We call 
$\mu_j(\lambda)$ the $N$ branches over $\lambda$. Using the Riemann--Hurwitz
formula, we can compute the genus of  $\Gamma$ \cite{BaTa99}:
$$
g={N(N-1)\over 2}\sum_k n_k -N +1 = {1\over 2} {\rm dim}\,{\cal M}
$$

It is important to observe that the
genus is  half the dimension of phase space.
So  the number of action variables occurring as
independent parameters in the eq.(\ref{specurve2}) should also
be equal to $g$.  Let us verify it. 
Since $r_j(\lambda)$ is the symmetrical function
$\sigma_j(\mu_1,\cdots,\mu_N)$, 
it is a  rational function of $\lambda $.
It has a pole of order $jn_k$ at
$\lambda =\lambda _k$.
Its value at $\lambda = \infty$ is known since
$\mu_j(\lambda)\to a_j$. Hence it can be expressed on $j\sum_k n_k$ 
parameters
namely the coefficients of all these poles.  Altogether we have
${1\over 2} N(N+1) \sum_k n_k$ parameters. They are not all independent
however. Above $\lambda =\lambda _k$ the various branches can be written:
\begin{equation}
\mu_j(\lambda)=\sum_{n=1}^{n_k} {c^{(j)}_n\over (\lambda -\lambda _k)^n}+{\rm
regular}
\label{cont1}
\end{equation} 
where all the coefficients $c^{(j)}_1,\cdots,c^{(j)}_{n_k}$
are fixed and non--dynamical because they are the matrix elements of
the diagonal matrices $(A_k)_-$, while the regular part is
dynamical. This implies on $r_j(\lambda)$ that the coefficients of its $n_k$ highest 
order pole terms  are fixed. Summing over $j$, we get
$Nn_k$ constraints and we are left with ${1\over 2} N(N-1) \sum_k n_k$ 
parameters, that is $g+N-1$ parameters.
It remains to take the quotient by the action of constant diagonal matrices.
The generators of this action are the Hamiltonians 
$H_n=(1/n)\,{\rm res}_{\lambda =\infty}{\rm Tr}
\,(L^n(\lambda ))\, d\lambda $,
i.e. the term in $1/\lambda $ in ${\rm Tr}\,(L^n(\lambda ))$. 
Setting 
\begin{equation}
\mu_j(\lambda) = a_j + {b_j \over \lambda} + \cdots 
\label{cont2}
\end{equation} 
around the point $Q_j = (\infty,a_j)$, we have $H_n = \sum_j a_j^{n-1} b_j $.
After Hamiltonian reduction these quantities are to be set to fixed
(non--dynamical) values. So, both $a_i$ (by definition) and $b_i$ are non dynamical. 
On the functions  $r_j(\lambda )$ this implies that their expansion at infinity 
starts as $r_j(\lambda ) = r_j^{(0)}+{ r_j^{(-1)}\over \lambda } + \cdots$,
with $r_j^{(0)}$ and $r_j^{(-1)}$ non dynamical.
Hence when the system is properly reduced we
are left with exactly $g$ action variables.

The constraints eqs.(\ref{cont1}, \ref{cont2}) can be summarized in a very elegant
way \cite{KrPh97, BaBeTa03}. 
Introduce the differential $\delta$ with respect to the dynamical moduli. 
Then our constraints mean that  the differential 
$\delta \mu d\lambda$ is regular everywhere on the spectral curve because the coefficients of the various poles being non dynamical, they are killed by $\delta$:
\begin{eqnarray*}
\delta \mu \;d\lambda =\;{\rm holomorphic}
\end{eqnarray*}
Since the space of holomorphic differentials is of dimension $g$, the right hand side
of the above equation is spanned by $g$ parameters which are the $g$ independent 
action variables we were looking for. Notice that these action variables are coefficients 
in the pole expansions of the functions $r_j(\lambda)$, and thus appear linearly 
in the equation of $\Gamma$. Hence eq.(\ref{specurve2}) can be written in the form
eq.(\ref{rlambdamu}). Clearly, these considerations can  be adapted  
by considering more general conditions such as
\begin{eqnarray*}
{\delta \mu \over \mu^n} \;{d\lambda\over \lambda^m} =\;{\rm holomorphic}
\end{eqnarray*}

\section{Examples.}
Let us show how well known models fit into our scheme. For the 
hyperelliptic ones, things are so simple that we can directly check
the commutation of the Hamiltonians.
\subsection{Neumann model.}
The spectral curve can be written in the form \cite{BaBeTa03}:
\begin{equation}
 \mu^2 = { \prod_{i=1}^{N-1}(\lambda-b_i) \over  \prod_{i=1}^{N}(\lambda-a_i) } 
 =  {P(\lambda) \over Q(\lambda) } 
\label{Neucurverational}
\end{equation}
Performing the birational transformation $ s= \mu Q(\lambda)$, we get:
\begin{equation}
s^2=Q(\lambda)P(\lambda)
\label{Neucurve3}
\end{equation}
which is an hyperelliptic curve of genus $g=N-1$. The polynomial $Q(\lambda)$
is non dynamical. We have $(N-1)$ independent
dynamical quantities, namely the $(N-1)$ symmetrical functions of the $b_i$, coefficients of 
$P(\lambda)$. 
We have
\begin{eqnarray*}
\delta \mu \;d\lambda =  {\delta P(\lambda) \over 2 \mu Q(\lambda)} d\lambda = 
 {\delta P(\lambda) \over 2 s} d\lambda =\;{\rm holomorphic}
\end{eqnarray*}
Asking that a curve of the form eq.(\ref{Neucurverational}) passes through
the $g$ points $(\lambda_i,\mu_i)$ determines the polynomial $P(\lambda)$.
The solution of Baxter's equations
$$
P(\lambda_i ) =  Q(\lambda_i) \mu_i^2
$$
simplifies in this case because the matrix $B$ depends only on the $\lambda_i$. 
It is equivalent to Lagrange interpolation formula:
$$
P(\lambda) = P^{(0)}(\lambda) + P^{(2)}(\lambda)
$$
with
$$
P^{(0)}(\lambda) =\prod_i (\lambda - \lambda_i), \quad 
P^{(2)}(\lambda) = \sum_j  S_j (\lambda) Q(\lambda_j)\mu_j^2,\quad
S_j(\lambda) = {\prod_{k\neq j}(\lambda - \lambda_j)\over 
\prod_{k \neq j} (\lambda_j - \lambda_k) } 
$$
Introducing the canonical commutation relations
$$
\relax [ \mu_j , \lambda_k ] = i \hbar \delta_{jk}
$$
so that
\begin{eqnarray*}
\relax [ \mu_j, f(\lambda_j) ] = i\hbar \partial_{\lambda_j} f(\lambda_j),\quad
\relax [ \mu_j^2, f(\lambda_j) ] = 2 i\hbar  \partial_{\lambda_j} f(\lambda_j) \mu_j 
+ (i\hbar)^2 \partial^2_{\lambda_j} f(\lambda_j)
\end{eqnarray*}
We can check that $ [ P(\lambda), P(\lambda') ] =0$ is a consequence of 
$ \partial_{\lambda_j}P^{(0)}(\lambda) = - \prod_{k\neq j} ( \lambda_j - \lambda_k)  S_j(\lambda)$, and the identities
$$
S_j(\lambda) \partial^n_{\lambda_j}S_i(\lambda') - S_j(\lambda') \partial^n_{\lambda_j}S_i(\lambda) =0, \quad \forall n >0
$$
These identities follow from the remark that, if we define the translation operators
$t_j \lambda_i = \lambda_j + \sigma \delta_{ij}$, then
\begin{equation}
 S_j(\lambda) t_j  S_i(\lambda') - S_j(\lambda') t_j  S_i(\lambda) = 
{ \prod_{k\neq i,j}(\lambda - \lambda_k) \prod_{k\neq i,j}(\lambda' - \lambda_k)
\over \prod_{k\neq j} (\lambda_j - \lambda_k) \prod_{k\neq i} (\lambda_i - \lambda_k) }
(\lambda_i - \lambda_j) (\lambda - \lambda')
\label{identitygene1}
\end{equation}
is independent of $\sigma$.

\subsection{Toda Chain.}

The spectral curve can be written in the form \cite{FaTa86, BaBeTa03}:
\begin{eqnarray}
\mu + \mu^{-1} = 2 P(\lambda) 
\label{gamma}
\end{eqnarray}
where  $2P(\lambda)=\lambda^{n+1} - \sum_{i=1}^{n+1} p_i \lambda^n+\cdots $ 
is a polynomial of degree
$(n+1)$.  The spectral curve is  hyperelliptic since it can
be written as
\begin{eqnarray}
s^2 = P^2(\lambda) - 1, \quad {\rm with}\quad  s = \mu -P(\lambda)
 \label{specbis} 
\end{eqnarray} 
The polynomial $P^2(\lambda)$
is of degree $2(n+1)$ so the genus of the curve is $g = n$. The number
of dynamical moduli is $g=n$ in the center of mass frame $\sum_{i=1}^{n+1} p_i  =0$.
We have
\begin{eqnarray*}
{\delta \mu \over \mu}\;d\lambda = {2 \delta P(\lambda) \over \mu - \mu^{-1}}  d\lambda = 
 {\delta P(\lambda) \over  s} d\lambda =\; {\rm holomorphic}
\end{eqnarray*}
Asking that the curve eq.(\ref{gamma}) passes through the $n$ points $(\lambda_i,\mu_i)$,
we get Baxter's equations. 
$$
 2 P(\lambda_i) = \mu_i + \mu_i^{-1} 
$$
Their solution is again given by
Lagrange interpolation formula:
$$
2P(\lambda) = P^{(0)}(\lambda) + P^{(1)}(\lambda)
$$
where
$$
P^{(0)}(\lambda) =  (\lambda + \sum_i \lambda_i) \prod_{i=1}^n (\lambda - \lambda_i), \quad
P^{(1)}(\lambda) =   \sum_i  S_i(\lambda) (\mu_i + \mu_i^{-1})
$$
The polynomial $P^{(0)}(\lambda)$ is of degree $n+1$, vanishes for $\lambda = \lambda_i$ and has no  $\lambda^n$ term.
Let the commutation relations of the separated variables be given by:
$$
\mu_j \lambda_j = q \lambda_j \mu_j , \quad \mu_j \lambda_i =  \lambda_i \mu_j,
\quad i \neq j
$$
Then again $[ P(\lambda), P(\lambda') ] =0$ as a result of eq.(\ref{identitygene1}),
where $t_j$ is interpreted as $t_j \lambda_j = q \lambda_j$, and the
facts that
\begin{eqnarray*}
 S_j(\lambda) t^{\pm 1}_j  P^{(0)}(\lambda')
- S_j(\lambda') t^{\pm 1}_j  P^{(0)}(\lambda) &=&    P^{(0)}(\lambda')  S_j(\lambda)  -  P^{(0)}(\lambda)   S_j(\lambda')  \\
&& \hskip -3cm
={ \prod_{k\neq j }(\lambda - \lambda_k )\prod_{k\neq j }(\lambda' - \lambda_k )
\over \prod_{k\neq j} (\lambda_j - \lambda_k) } (\lambda + \lambda'
+ \sum_{i\neq j} \lambda_i)(\lambda'-\lambda)
\end{eqnarray*}

\subsection{A non--hyperelliptic model.}
We consider the model studied in \cite{Skly92, Smi01,SmiZe02} . 
The spectral curve can be written in the form:
\begin{equation}
R(\lambda,\mu) \equiv \mu^N + t^{(1)}(\lambda) \mu^{N-1} + \cdots t^{(N)}(\lambda) = 0
\label{FedVacurve}
\end{equation}

The polynomials $t^{(k)}(\lambda)$ are such that 
${\rm degree}\; t^{(k)}(\lambda) \leq kn -1$ and 
$ {\rm degree}\; t^{(N)}(\lambda) = Nn -1$ 
for some integer $n$. The genus of this curve is
$$
g = {1\over 2} (N-1)(Nn-2)
$$
Assuming that there is no singular point at finite distance, the homomorphic differentials are
$$
\omega_{kl} = { \mu^l \lambda^k \over \partial_\mu R(\lambda,\mu) } d\lambda,\quad
0 \leq l < N-1,\quad 0 \leq k < (N-l-1)n -1
$$
We have
$$
\delta \mu {d\lambda \over \lambda} = - \sum_{k=1}^N {\delta t^{(k)}(\lambda) \mu^{N-k} 
\over \partial_\mu R(\lambda, \mu) } {d\lambda \over \lambda}
$$
This will be holomorphic if $\delta t^{(1)}(\lambda) = 0$ and
$$
\delta t^{(k)}(\lambda) =\delta H^{(k)}_1 \lambda  + \cdots  + \delta H^{(k)}_{(k-1)n -1}\lambda^{(k-1)n -1}
$$
Baxter's equations and the commutation of the Hamiltonians
where proved in this case, starting from the definition of the quantum model
through it Lax matrix and the Yang--Baxter equation. Our approach gives a very simple 
proof of this result.

\section{Conclusion.}

We have shown that starting from the separated 
variables, one can give an easy definition of a quantum integrable system.
The next step is to reconstruct the Lax matrix and the original dynamical
variables of the model. While this is a well understood problem in the classical theory 
\cite{DuKrNo90, BaBeTa03},
(it is the essence of the classical inverse scattering method), its quantum 
counterpart will require a deeper understanding of the quantum affine Jacobian \cite{Smi00}.

{\bf Acknowledgements.} We thank D. Bernard and F. Smirnov for discussions.

 \end{document}